\documentclass[pra, a4paper, twocolumn, 10pt, showpacs]{revtex4}
\usepackage{bm, bbm, amsmath, amssymb, amsthm}
\usepackage[T1]{fontenc}
\usepackage[latin9]{inputenc}
\usepackage{graphicx}
\usepackage{geometry}
\newcommand{\be}{\begin{eqnarray}}
\newcommand{\ee}{\end{eqnarray}}

\newcommand{\ket}[1]{\left|{#1}\right\rangle}
\newcommand{\bra}[1]{\left\langle{#1}\right|}
\newcommand{\braket}[2]{\langle{#1}|{#2}\rangle}

\begin{document}

\title{Concatenated tensor network states}

\author{R.\ Hübener$^{1}$, V.\ Nebendahl$^{1}$, and W. Dür$^{1,2}$ }

\affiliation{$^1$ Institut für Theoretische Physik, Universität
Innsbruck, Technikerstraße 25, A-6020 Innsbruck,
Austria\\
$^2$ Institut für Quantenoptik und Quanteninformation der Österreichischen Akademie der Wissenschaften, Innsbruck, Austria}

\date{\today}

\begin{abstract}
We introduce the concept of concatenated tensor networks to efficiently describe quantum states. We show that the corresponding concatenated tensor network states can efficiently describe time evolution and possess arbitrary block-wise entanglement and long-ranged correlations. We illustrate the approach for the enhancement of matrix product states, i.e. 1D tensor networks, where we replace each of the matrices of the original matrix product state with another 1D tensor network. This procedure yields a 2D tensor network, which includes -- already for tensor dimension two -- all states that can be prepared by circuits of polynomially many (possibly non-unitary) two-qubit quantum operations, as well as states resulting from time evolution with respect to Hamiltonians with short-ranged interactions. We investigate the possibility to efficiently extract information from these states, which serves as the basic step in a variational optimization procedure. To this aim we utilize known exact and approximate methods for 2D tensor networks and demonstrate some improvements thereof, which are also applicable e.g. in the context of 2D projected entangled pair states. We generalize the approach to higher dimensional- and tree tensor networks.
\end{abstract}


\maketitle

\tableofcontents{}


\section{Introduction}\label{sec:intro}

The classical simulation of complex quantum systems is one of the central problems in modern physics. Given the exponential growth of the state space with the system size, such a classical simulation seems infeasible. However, it has been realized that quantum systems occurring in nature often do only populate a small subspace. Identifying this subspace is hence the first step towards a successful classical simulation. For ground states of (non-critical) strongly correlated quantum spins in a one dimensional setup, matrix-product states (MPS) \cite{FNW92, AKLT88, VPC04} turn out to provide a proper parametrization for this subspace \cite{Ha07, VC06}. MPS can not only efficiently describe such ground states, but it is also possible to efficiently read out physical information from this description, e.g., to compute expectation values of local observables and correlation functions. Moreover, MPS form the basis of the density matrix renormalization group (DMRG) \cite{Wh91, Sc04}, a powerful numerical method that has been successfully applied to various problems in 1D. The relation between the DMRG and MPS is an example how physical insight into the logic of a preparation (renormalization) procedure can be manifestly encoded into the structure of a state class.

Recent approaches to simulate ground states of strongly correlated systems in critical systems or higher dimensions follow a similar approach. A variety of states such as projected entangled pairs (PEPS) \cite{VC06b}, sequentially generated states \cite{CB08}, string-bond states \cite{SWV07}, weighted graph states \cite{ABD06, ABD07}, renormalization ansatz with graph enhancement \cite{HKHDVEP08} or the multiscale entanglement renormalization ansatz (MERA) \cite{Vi07} have been introduced with the aim of efficiently parametrizing the relevant subspace. The entanglement properties of the corresponding states form the guideline and determine the potential applicability of the methods. For instance, MERA can provide a logarithmic divergence for block-wise entanglement in critical 1D systems, while e.g.\ 2D variants of MERA as well as PEPS and string-bond states fulfill area laws for block-wise entanglement, typically to be found in ground states of 2D systems. In all cases it is crucial that not only an efficient description of the states can be obtained, but also that information can be efficiently extracted, either exactly or in an approximate way. Based on these states, variational methods for ground state approximation and (real and imaginary) time evolution have been developed and tested. While MPS, MERA and PEPS lead to good descriptions of ground states for non-critical 1D systems, critical 1D systems and 2D systems respectively, none of the proposed classes seems to be suitable to properly describe time evolution. In fact, it has been argued that simulating time evolution is in general hard \cite{SWVC08}, as the block-wise entanglement grows -- already for 1D systems -- linearly in time, leading quickly to a volume law. The entanglement contained in an MPS is bounded by the dimension of the matrices or tensors, and the entanglement contained in a PEPS follows an area law.

Here we present a class of tensor network states for which such limitations do not apply, and which allow one in principle to efficiently describe states resulting from time evolution or quantum computation. To construct these states we make use of the basic idea underlying previous tensor network structures. In these structures, a simplification of the existing description can be achieved by replacing tensors of high rank (i.e., with many indices) by a \emph{network} of tensors of low rank (i.e., with few indices) with appropriate topology. The choice of the underlying tensor network determines qualitatively different sub-classes of states, in previous approaches e.g.\ having lead to MPS or PEPS when describing a 1D or 2D structure respectively. We apply this idea in an iterative, or {\em concatenated}, fashion, leading to concatenated tensor network states (CTS). That is, each of the tensors appearing in a tensor network is itself repeatedly replaced by another tensor network. The resulting structure is again a tensor network, similar to a PEPS, with the main difference that only some of the tensors are associated with physical particles.

The efficient and exact extraction of information, e.g., expectation values or correlation functions, from an arbitrary tensor network is in general not possible, as they rely on a contraction of the network, i.e., summations over all indices of the network. Even for 2D tensor networks, the contraction is known to be computationally hard (\#P-hard) \cite{Ba82}. However, for certain special cases exact evaluation is possible. In addition, also \emph{approximate} contraction and certain Monte-Carlo methods have been developed and successfully applied in the context of 2D PEPS and imaginary time evolution \cite{MVC07}. We demonstrate the applicability of the established methods to the CTS and several enhancements thereof. We moreover demonstrate that there are novel implementations of algorithms like (imaginary) time evolution of 1D systems and the application of quantum circuits that are more efficient in the CTS than in MPS.

This paper is structured as follows. In Sec.~\ref{sec:cts}, we will introduce the CTS, give examples and illustrate their properties from an analytic point of view. In Sec.~ \ref{sec:properties}, we discuss the applications of CTS and illustrate the potential of the CTS to describe states relevant in physics. As an example, we give the numerical treatment of a toy model, more precisely, we will describe a state originating from the time evolution of a product state governed by the Ising Hamiltonian. In Sec.~ \ref{sec:applications} we finally show several ways to extract information from a CTS, thereby utilizing and improving methods to (approximately) contract 2D tensor networks.


\section{Concatenated tensor networks}\label{sec:cts}

In this section we introduce the CTS and in the context of the problems having lead to tensor network descriptions in general.

A generic quantum state of $N$ $d$-level systems can be written in a basis whose elements are tensor products of basis states of the local $d$-level systems. The quantum state is then characterized by the coefficients of these basis states, which are tensors $A_{s_1s_2\ldots s_N}$ of rank $N$ and dimension $d$
\be
|\psi\rangle= \sum_{s_1,s_2,\ldots,s_n=1}^d A_{s_1s_2\ldots s_N} |s_1s_2\ldots s_N\rangle \label{psi}.
\ee
Hence, the description of such a state consists of $d^N$ complex parameters. This exponential growth of the number of parameters used in the generic description makes it unsuitable for numerical analysis. 


\subsection{MPS and PEPS}

The tensor $A_{s_1s_2\ldots s_N}$ of the generic description given above can be decomposed into a tensor network, thereby imposing a structure in this set of parameters. To do so, we will represent $A_{s_1s_2\ldots s_N}$ by another set of tensors of smaller rank. Some of the indices of the small-rank tensors correspond to the state of a physical site $\{s_1,s_2\ldots,s_N\}$ as before. The remaining auxiliary indices are shared between pairs of the small-rank tensors, and to recover the coefficient of a basis state of the physical system, the shared indices will be contracted, i.e., summed over. The information which tensors share indices can be represented by a graph, where ``tensors'' correspond to vertices and ``sharing an index'' corresponds to an edge. The indices corresponding to physical states will in the following be called open. We will furthermore use greek letters $\alpha_j,\beta_k$ etc.\ to refer to shared indices, while open indices will be denoted by $s_j$.

\begin{figure}[ht]
\includegraphics[width=.45\textwidth]{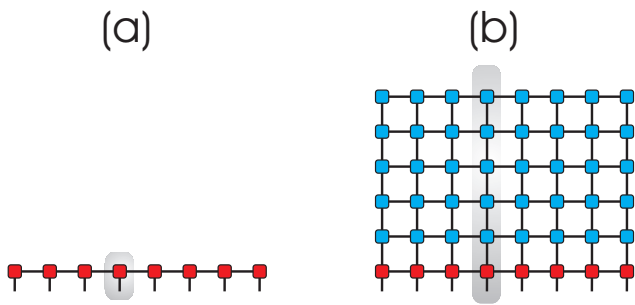}
\caption{(Color online). (a) Graphical representation of a 1D tensor network (MPS). The boxes correspond to tensors, where shared indices are summed over. Open indices correspond to physical particles (red tensors). (b) Each of the tensors in the original tensor network is replaced by a 1D tensor network (matrix product operator) arranged in $y$-direction. Auxiliary tensors (no open indices) are drawn in blue. This leads to a 2D tensor network.}
\label{Fig_MPS_concat}
\end{figure}

As an example, one may use a 1D structure for the tensor network, leading to MPS (see Fig.~\ref{Fig_MPS_concat}a)
\begin{multline}
A_{s_1s_2\ldots s_N}  \\
= \sum_{\alpha_1,\alpha_2,\ldots \alpha_N=1}^D A_{\alpha_1}^{[s_1]}A_{\alpha_2\alpha_3}^{[s_2]}\ldots A_{\alpha_{N-1}\alpha_N}^{[s_{N-1}]}A_{\alpha_N}^{[s_N]},
\end{multline}
which are described by the tensors $A_{\alpha_i\alpha_{i+1}}^{[s_i]}$ and $A_{\alpha_1}^{[s_1]}, A_{\alpha_N}^{[s_N]}$. For a fixed choice of $s_1s_2 \ldots s_N$, the coefficient $A_{s_1s_2\ldots s_N}$ is obtained by calculating the product of the $D \times D$ matrices $A_{\alpha_i\alpha_{i+1}}^{[s_i]}$ (except at the border, where one has vectors $A_{\alpha_i}^{[s_i]}$). By choosing $D$ large enough (but still $D \leq d^N$), one can represent any tensor and hence any quantum state in this form. A restriction to small $D$ allows to describe a certain subset of states efficiently.

\begin{figure}[ht]
\includegraphics[width=.45\textwidth]{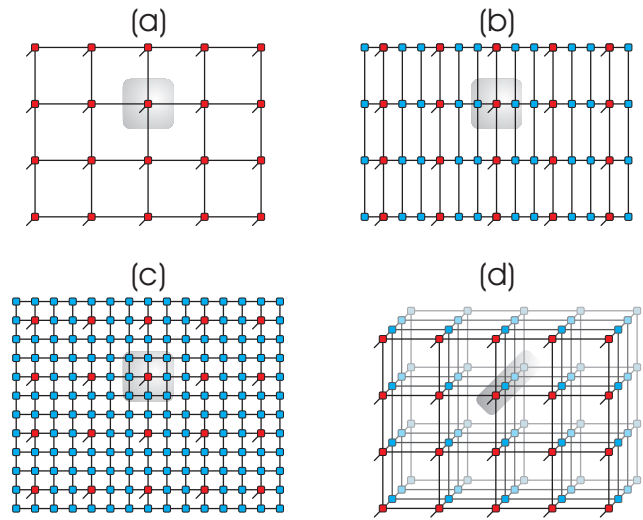}
\caption{(Color online). Examples of concatenated 2D tensor network states. The boxes correspond to tensors, where joint indices are summed over. Open indices correspond to physical particles (red tensors), while auxiliary tensors (no open indices) are drawn in blue. (a) Original 2D tensor network,  where each of the tensors corresponds to a physical particle. (b) Each of the original tensors is replaced by a 1D tensor network (MPS, consisting of 3 tensors, two of which are auxiliary tensors) in horizontal direction. (c) Each of the original tensors is replaced by a 2D tensor network (of size $3\times3$) arranged in the same plane as the original 2D tensor network.  (d) Each of the initial tensors is replaced by an MPS perpendicular  to the original plane ($z$-direction). This leads a 3D tensor network structure.}
\label{PEPS_2D_concat}
\end{figure}

In a similar way, one can consider tensor network structures with different topology and higher dimensional connectivity. If the physical system consists of particles on a 2D regular lattice, the procedure analogous to the construction of the MPS described above yields a 2D regular tensor grid, e.g.,
\begin{multline}
A_{s_1s_2\ldots s_9}= \sum_{\mathrm{greek\, indices}=1}^D A_{\alpha_1,\beta_1}^{[s_1]}
A_{\alpha_1\alpha_2\beta_2}^{[s_2]}
A_{\alpha_2\beta_3}^{[s_3]}
A_{\beta_1\alpha_3\beta_4}^{[s_4]}\\ \notag
\times A_{\alpha_3\beta_2\alpha_4\beta_5}^{[s_5]}
A_{\beta_3\alpha_4\beta_6}^{[s_6]}
A_{\beta_4\alpha_5}^{[s_7]}
A_{\alpha_5\beta_5\alpha_6}^{[s_8]}
A_{\alpha_6\beta_6}^{[s_9]}
\end{multline}
the structure corresponding to the projected entangled pair states (PEPS) in 2D (see Fig.~\ref{PEPS_2D_concat}a).

Tensor networks that have been subject to detailed investigation include one-dimensional graphs with and without periodic boundary conditions (MPS), trees \cite{SDV06, MS06} and two-dimensional lattices (PEPS) \cite{VC06b}. Investigations of networks of different topology have shown that 1D and tree-like structures are generally easy to simulate numerically. Tensor networks corresponding to graphs with many loops, on the other hand, are generally hard to simulate \cite{SDV06,MS06} and only in special cases efficient algorithms are known, see, e.g., Ref.~\cite{MGcomb}. Some of the networks, e.g. corresponding to a 2D lattice, are even known to correspond to states being resources of measurement based quantum computation and hence (having a generally applicable method) to treat these tensor network states numerically efficiently and exactly would mean to efficiently simulate a quantum computer classically. In fact, the contraction of such 2D networks was proven to be a computationally hard problem in general \cite{SWVC07}.


\subsection{Concatenated tensor network states}

We will consider \emph{concatenated tensor networks} in the following. That is, given a tensor network as in the previous subsections, we will replace each individual tensor in the network by another tensor network. This can in principle be done in an iterative way, leading to concatenated tensor structures. We will typically only consider tensor networks stemming from few iterations, given the fact that the total number of tensors increases exponentially with the number of iterations. Notice that most of the additional tensors that we introduce will be auxiliary tensors, i.e., without open indices and hence not corresponding to quantum systems. We also remark that it is not necessary to use the same tensor structure at each concatenation level.

The structure that we finally obtain is again a (possibly high-dimensional) tensor network. As long as the total number of tensors, as well as their rank and dimension, is polynomially bounded, we obtain a class of states that can be described by a polynomial number of parameters, i.e., efficiently. We call the family of quantum states that can be described in this way \emph{concatenated tensor networks states} (CTS).

The key element of this approach is to impose internal structure on the tensor description being used, thereby reducing the information content, while its ability to describe entanglement is in principle kept. This allows one to describe states with a large amount of block-wise entanglement, up to a volume law, and long-ranged correlations using only small rank tensors of small dimension at the elementary level.

By construction, the imposed structure is similar to the one behind the very successful DMRG renormalization. Independently of this ancestry of the ansatz, there are some illuminating interpretations going beyond the DMRG picture. Different from the DMRG, the renormalization structure in the concatenated tensor network is not necessarily applied in a \emph{spacial} fashion, but (being subject to interpretation and depending on the actual network) in a timely fashion, e.g., as preparatory applications of certain operators in a Suzuki-Trotter expansion, or going further, as state-preparing applications of generic operators. We find that already with a 2D network of size ${\rm poly}(N)$ and tensor dimension two, \emph{all} states that can be prepared by a polynomially sized quantum circuit can be represented as CTS. Furthermore, the picture that the CTS stems from a preparation using measurement based quantum computation (MQC) is possible. All these interpretations are suited to inspire further development and nurture some hope that the described state class might -- by virtue of its construction -- be suited for a good description of time evolved states or quantum circuits.


\section{Properties of concatenated tensor network states}\label{sec:properties}

In the following, we give a number of examples of CTS and discuss their properties.


\subsection{Concatenated MPS}\label{MPS_concat}

We now consider concatenated MPS. We start with a 1D tensor network as shown in Fig.~\ref{Fig_MPS_concat}a, and replace each of the tensors $A_{\alpha_k\alpha_{k+1}}^{[s_k]}$ by a 1D tensor network, as shown in Fig.~\ref{Fig_MPS_concat}b. More precisely, each matrix $A_{\alpha_k\alpha_{k+1}}^{[s_k]}$ for $s_k=1,2,\ldots,d$ is replaced by a matrix product operator (MPO) \cite{MCPV08},
\be
&&A_{\alpha_k\alpha_{k+1}}^{[s_k]} \leftrightarrow  \\
&&\sum_{\beta_1,\beta_2,\ldots \beta_M=1}^{D_k}
A_{\alpha_k^1\alpha_{k+1}^1\beta_1}^{[s_k]}
B_{\alpha_k^2\alpha_{k+1}^2\beta_1\beta_2}\ldots
B_{\alpha_k^M\alpha_{k+1}^M\beta_M}.\nonumber
\ee
and the indices $\alpha_k$ are replaced by $\alpha_k^{j} \in (1,2,\ldots,D)$ corresponding to several connections to the neighboring tensors. Notice that the effective dimension of all these connections together is given by $\chi=\prod_k D_k$. In this way we obtain a 2D tensor network, where only $N$ tensors $A^{[s_k]}$ have open indices and correspond to physical sites, while there are $(N-1)M$ auxiliary tensors $B$. The process of replacing individual tensors by 1D tensor networks can be iterated. At the next level, one obtains a 3D tensor network and so forth. We remark that one may also consider 2D tensor networks with periodic boundary conditions, either in horizontal or vertical direction.

In the following we will consider a 2D tensor networks (i.e., only the first iteration level) of size $N \times M$ with $M= {\rm poly}(N)$. We analyze the states that can be described by such a CTS, and study their entanglement features. We show that
\begin{itemize}
\item All states that can be created by a polynomially sized quantum circuit can be efficiently described by such a 2D CTS with $D_k=2$. This includes {\em unitary} quantum circuits as well as {\em post-selected} quantum circuits.
\item All states resulting from a time evolution for a time $t$ with respect to short-range Hamiltonians can be efficiently described by an $N \times M$ 2D CTS, where $M$ scales quadratically with time $t$.
\item A subclass of matrix product states with an effective bond-dimension of the order of $\chi=D_k^M$ can be described efficiently by an $N \times M$ 2D CTS.
\end{itemize}
Regarding the entanglement features, we show
\begin{itemize}
\item The block-wise entanglement of an $N \times M$ 2D CTS can be $O(M)$. In particular, states with a volume law for block-wise entanglement and with long-ranged correlations can be described efficiently.
\end{itemize}


\subsubsection{Interpretation in terms of (post selected) quantum circuits}\label{MPS_circuit}

Here we show that for a specific choice of tensors the 2D tensor network can be interpreted as a quantum circuit consisting of generic gates. We consider a quantum circuit for $N$ qubits of depth $M=O({\rm poly}(N))$. We find that one can describe the resulting state from such a quantum computation by a 2D tensor network of size of order $O(N \times M)$, i.e., of polynomially many tensors, where the tensor dimension is $D=2$. Let us now demonstrate how a standard quantum circuit consisting of arbitrary single-qubit rotations and two-qubit phase gates -- which constitute a universal gate set -- can be encoded into the tensor network.
We denote the auxiliary tensors by $B_{\alpha_l\alpha_r\alpha_u\alpha_d}^ {(i,j)}$ and the ones connected to physical particles by $A_{\alpha_l\alpha_r\alpha_u s_d}^ {(i,j)}$ (typically), where the sub-indices $l,r,u,d$ stand for left, right, up and down, and $i,j$ are labels that indicate the position of the tensor in the 2D tensor network ($i^{\rm th}$ row and $j^{\rm th}$ column). The uppermost line of tensors $B_{\alpha_l\alpha_r\alpha_d}^ {(1,j)}$ have no "up" index, and similarly the tensors at the border do not have left/right indices. We identify each horizontal line of tensors with a certain time step in the circuit, and the first (uppermost) line is used to initialize the input state to $|0\rangle^{\otimes N}$ (or some other product state), while the last line corresponds to the output state.

Initialization can, e.g., be achieved by choosing $B^{(1,j)}_{000}=1$ and all other entries $0$, where we identify the component $0$ (1) of the down link with the state $|0\rangle$ ($|1\rangle$). The basic idea is then to either erase the left-right links between two neighboring tensors, so that processing of individual qubits can be performed, or make use of this link to perform an (entangling) two qubit gate. In the contraction of the tensor network, one sums over all possible values for each of the links. Hence if we choose $\forall \alpha_r\alpha_u\alpha_d : B_{0\alpha_r\alpha_u\alpha_d} ^{(i,j)} = 0$, the link to the left is essentially broken~\footnote{This is not the only possibility to break a link. Also the choice $B_{0\alpha_r\alpha_u\alpha_d}^{(i,j)} = B_{1\alpha_r\alpha_u\alpha_d}^{(i,j)}$ is possible if also neighboring tensors are constructed in the same way.}. Similarly, the link to the right can be broken by choosing $\forall \alpha_l\alpha_u\alpha_d : B_{\alpha_l0\alpha_u\alpha_d}{(i,j)} = 0$.
Hence the choice
\be
\label{SU2}
B_{11\alpha_u\alpha_d}^{(i,j)}=U_{\alpha_u\alpha_d}
\ee
(and all other entries are 0) allows us to implement the single-qubit (unitary) operation
\be
U=\sum_{\alpha_{d},\alpha_u=0}^1 U_{\alpha_d\alpha_u}|\alpha_d\rangle\langle\alpha_u|
\ee
on qubit $j$ in time step $i$.

For a two-qubit phase gate diag$([1,1,1,-1])$, i.e.,
\be
U_{\rm PG}&=&\sum_{\alpha_d,\beta_d,\alpha_u,\beta_u =0}^1 U_{\alpha_d\beta_d\alpha_u\beta_u} |\alpha_d \beta_d\rangle \langle \alpha_u\beta_u| \nonumber\\
&=& \sum_{\alpha_d,\beta_d=0}^1 (-1)^{\alpha_d \cdot \beta_d} |\alpha_d \beta_d\rangle \langle \alpha_d\beta_d|,
\ee
acting on qubits $j,j+1$ in time step $i$, we find that the following choice of tensors allows one implement this gate: $B_{1000}^{(i,j)}= B_{1011}^{(i,j)}= B_{1111}^{(i,j)}=1$; $B_{0100}^{(i,j+1)}= B_{0111}^{(i,j+1)}=1, B_{1111}^{(i,j+1)}=-2$, while all other tensors are zero. This can be seen by noting that the links to left (particle $j-1$) and right (particle $j+1$) are broken, and by contracting the two tensors over their joint index $(\alpha_r,\beta_l)$. Other two qubit gates corresponding to the class of CNOT and phase gates \cite{DC02} (i.e., gates that can create only Schmidt-rank two states or only one e-bit of entanglement) can be realized. Among these gates are e.g. controlled phase gates with a controllable phase $\varphi$, $U_{\rm PG}(\varphi)= diag([1,1,1,e^{i\varphi}])$.

To give an example for a subclass of states with a large amount of entanglement to be created by operators and to be hold by a simple CTS description, consider controlled phase gates $U_{\rm PG}(\varphi)$ between arbitrary pairs of particles initially prepared in $|+\rangle=1/\sqrt{2} (|0\rangle + |1\rangle)$. These circuits prepare weighted graph states (WGS) \cite{ABD06, ABD07}, utilizing only $O(N^2)$ gates. Using nearest neighbor gates, one needs at most $O(N^3)$ phase gates to prepare an arbitrary WGS, although one is not restricted to these in our setup. As demonstrated in \cite{ABD06}, WGS can have maximal block-wise entanglement, maximal localizable entanglement as well as long-ranged correlations. Similarly, as shown in \cite{ODP07}, typical states with $O(L)$ block-wise entanglement for all blocks of length $L$ can be generated by $O(N^3)$ two-qubit gates acting on arbitrary pairs of particles, leading to a tensor network of size $N \times O(N^4)$. 

The generalization to other (non-unitary) circuits or other elementary gates is straightforward. For instance, the unitary matrix $U_{\alpha_d\alpha_u}$ in Eq.~\ref{SU2} can be replaced by an {\em arbitrary} matrix $A_{\alpha_d\alpha_u}$, corresponding to an arbitrary single-qubit operation. In particular, a single-qubit measurement {\em with a selected outcome} can be described in this way by choosing $A$ to be a 1D projector. Using such a construction, one obtains all states that can be described by an arbitrary post-selected quantum circuit. The corresponding complexity class is postBQP, which is in fact equivalent to PP~\cite{Zoo}.

Finally, we remark that, when considering a 2D tensor network on a tilted lattice, one can interpret the tensors directly as (unitary or non-unitary) quantum gates acting on nearest neighbors (see also Ref.~\cite{VDRB}).


\subsubsection{Description of time evolution}\label{MPS_time}

Similarly to the description of a polynomially sized quantum circuit, one can find, as a special case, a description of time evolution in terms of a polynomially sized 2D tensor network. Consider for example a nearest-neighbor 1D Hamiltonian $H=\sum_j {H_{j,j+1}}$ that we decompose into two parts, $H_1$ and $H_2$, where $H_1$ [$H_2$] contains pairwise commuting terms acting on different systems. That is, $H_1=\sum_k H_{2k-1,2k}$, while $H_2=\sum_k H_{2k,2k+1}$, see Refs.~\cite{Vi03,Vi04}. Using the Suzuki-Trotter expansion, we can write
\be
e^{-itH} &=& e^{i(H_1+H_2)t} \notag \\
&=& \lim_{n\to \infty} \prod_{k=1}^{n}(e^{-i H_1 t/n}e^{-i H_2 t/n}), \notag
\ee
where for a fixed time $t$ we obtain a proper approximation with bounded error $\epsilon$ by choosing $n=O(t^2/\epsilon)$, see Ref.~\cite{JVDZC03}, and hence a fixed small time step $\delta t = t/n = O(\epsilon/t)$. Hence the time evolution for time $t$ is accurately described by a sequence of $2n$ gates of the form $e^{-i\delta t H_j}$, where $n$ scales quadratically with~$t$~\cite{SS99}. Each of the gates $e^{-\delta t H_j}$, $j =1,2$ can be described by a 2D tensor network of size $N \times c$, where $c$ is a small constant, similarly as discussed for polynomially sized quantum circuits in the previous subsection. The state resulting from a time evolution for time $t$ with respect to the Hamiltonian $H$ applied to some initial product state can hence be described by a 2D tensor network of size $N \times M$ with $M=2cn = O(t^2/\epsilon)$.


\subsubsection{Interpretation in terms of measurement-based quantum computation.}

Another interpretation of such a tensor network description is provided by measurement-based quantum computation (MQC) \cite{Ra01,RB01}. One can view the 2D tensor network as the PEPS description of e.g. a 2D cluster state, where all but $N$ particles (last row) are measured out. The choice of tensors allows one to choose the measurement directions of the corresponding (auxiliary) particles. In turn, the measurement pattern (i.e., the choice of measurements) determines the quantum state that is generated at the output qubits (corresponding to the open legs in our tensor network). In fact, as each choice of tensor corresponds to a specific measurement outcome, we consider only a {\em single} branch of the measurement-based quantum computation, i.e. probabilistic MQC with some non-zero success probability \cite{MPDVMBprep}. Again, this is equivalent to all post-selected quantum circuits. Notice that also other tensor structures are universal in this probabilistic sense \cite{VMDBcomb}, i.e., allow one to describe/generate all quantum states.

In other words, the tensor network describes a quantum state of $N+M$ particles, where the $M$ auxiliary particles are measured out in order to finally generate a state of $N$ quantum particles. The auxiliary particles (auxiliary tensors) allow one to assist the generation of an enlarged class of states.


\subsubsection{Interpretation as MPS with large effective dimension}

A general MPS corresponding to a 1D tensor network with matrix dimension $\chi$ is described by $N O(\chi^2)$ parameters. The block-wise entanglement in such a MPS is limited by $\log_2 \chi$. For a 2D CTS of size $N \times M$, and tensors of dimension $D$, we observe that one may still interpret the resulting state as an MPS or 1D network (by contracting the MPO along the vertical direction). The effective matrix dimension of the corresponding MPS is now given by $\chi=D^M$. This also implies that the potential block-wise entanglement, measured by the entropy, between systems $(1\ldots k)$ and $(k-1 \ldots N)$  is given by $\log_2 D^M = M \log_2 D$. This corresponds to an exponential increase in effective bond-dimension while increasing the total number of parameters to describe the state only polynomially. Clearly, only a specific subset of states with a given block-wise entanglement can be described by such a 2D tensor network, however this set now includes states with large block-wise entanglement. If $M = O(N)$, it follows that the corresponding states can even be maximally entangled, i.e., fulfill a volume law.

Notice that describing states in terms of such a 2D CTS can already be useful for small $M$. Consider for instance ground states of 1D critical systems, where it is known that a good description in terms of an MPS requires a matrix dimension $\chi = O(2^{\rm log N})$ \cite{VC06, Ha07}. Similarly, the states resulting from a time evolution for a time $t$ with respect to a nearest-neighbor Hamiltonian possess block-wise entanglement growing linearly with $t$, leading eventually to volume laws. This implies that a description in terms of a general MPS requires matrices of dimension $\chi=O(2^N)$, i.e., exponentially many parameters. In turn, the 2D CTS can possess block-wise entanglement scaling as $O(M \log D)$, while the total number of parameters is of order $O(MND^4)$. That is, already for $D$ fixed and $M = O(N)$ a volume law can be obtained. For a specific example for the successful application of such a CTS description in the context of time evolution, see Sec.~\ref{sec:applications}.

The natural limitation of the entanglement we describe is not given by its \emph{quantity}, which can be expressed, for example, as the cardinality of the set of Schmidt coefficients in a bipartition of the given state. The limitation underlying the efficiency is rather introduced by a certain \emph{structure}, or order, within this (potentially very large) set of Schmidt coefficients. Depending on the situation, the (itself variable) structure of the entanglement will not have such a big impact on accuracy that the limitation of the quantity would have.


\subsection{Concatenated PEPS}\label{2DPEPS}

We now turn to (the CTS extension of) 2D tensor networks of size $N\times N$, or equivalently 2D PEPS. In contrast to 2D networks considered in the previous subsection, all tensors in such a 2D tensor network have open indices and are hence associated with a physical system. As before, we now replace each of these tensors $A_{\alpha_l\alpha_r\alpha_u\alpha_d}^{[s_{i,i}]}$ by another tensor network. There are several possibilities to do this (see Fig.~\ref{PEPS_2D_concat}),
\begin{enumerate}
\item[(i)] We use a 1D tensor network (matrix-product operator) of dimension $D$ with $M$ tensors, arranged in horizontal direction. One of the tensors has an open index corresponding to a physical system, while $M-1$ are auxiliary tensors. This leads to a $(NM) \times N$ 2D tensor network depicted in Fig.~\ref{PEPS_2D_concat}b. Similar, one can use a 1D network arranged in vertical direction, leading to a $N \times (NM)$ 2D network.
\item[(ii)] We use a 2D tensor network of size $M \times M$ and dimension $D$, arranged in the same plane as the initial 2D network. One of the tensors has an open index corresponding to a physical system, while $M^2-1$ are auxiliary tensors. This leads to a $(NM) \times (NM)$ 2D tensor network depicted in Fig.~\ref{PEPS_2D_concat}c.
\item[(iii)] We use a 1D or 2D tensor network (see (i),(ii)), but arranged perpendicular to the initial 2D plane. This leads to a 3D tensor network as shown in Fig.~\ref{PEPS_2D_concat}d.
\end{enumerate}

In each case, one may apply the method in an iterated fashion. For simplicity, we will consider only the networks at the first iteration. Similar as in the case of concatenated MPS, the states resulting in (i) can be interpreted as a 2D tensor network, but with increased (virtual) dimension $\chi=D^M$ in either horizontal or vertical direction. Similarly, in (ii) we obtain states corresponding to a 2D tensor network with a virtual dimension $\chi=D^M$ in horizontal and vertical direction.

Note that already for very small $M$, the resulting states are useful, e.g., for a better approximation of ground states in 2D systems or to simulate time evolution in 2D. The advantage is that, while the underlying tensor structure is still two-dimensional -- reflecting the geometry of a 2D system -- one obtains with a relatively small overhead (a factor of $M^2$) an exponential increase of the (virtual) tensor dimension, $\chi_{\rm eff}=D^M$. Given the fact that variational methods based on 2D tensor networks show a rather unfavorable scaling with the tensor dimension ($O(D^{12})$ for computational cost and $O(D^8)$ for memory \cite{MVC07}), one may use this approach to achieve virtual large tensor dimensions while keeping the dimension of the elementary tensors -- and hence the computational cost and required memory -- small. Although there is also an increase of the computational cost with the total number of tensors (scaling as $O(N^2)$ \cite{MVC07}); it is however much more favorable. Neglecting effects such as additional sweeps for optimization, the computational effort is increased by a factor of $O(M^2)$. For instance, if $M=3$,$D=2$, we obtain an additional computational overhead of roughly one order of magnitude due to larger number of tensors, while the virtual tensor dimension is now given by $8$. Using the initial $N \times N$ network with tensor dimension $8$ would lead to a computational overhead factor of about $10^7$ as compared to the $D=2$ case. This difference becomes even more drastic when considering larger $D$ or $M$.

The approach (iii) is the analogue of concatenated MPS we considered in the previous subsection. Similar as for concatenated MPS, one can describe all states resulting from a (post-selected) polynomially sized quantum computation in this way if $M={\rm poly}(N)$. When considering a 3D tensor network as in (iii), one obtains single-qubit gates as well as nearest-neighbor gates acting on particles arranged on a rectangular 2D lattice. That is, the $z$-axis corresponds to the time axis, and the $x-y$ plane corresponds to state of the $N \times N$ particles arranged on the 2D lattice after applying the (post) selected quantum circuit. Similar to concatenated MPS, also the interpretation in terms of a time evolution (of particles on a 2D array with nearest-neighbor couplings) is possible. Moreover, one may use the 3D structure as ansatz states for a variational method to describe ground states or time-evolved states corresponding to some 2D systems.


\subsection{Tensor tree states with internal structure}

\begin{figure}[ht]
\includegraphics[width=.40\textwidth]{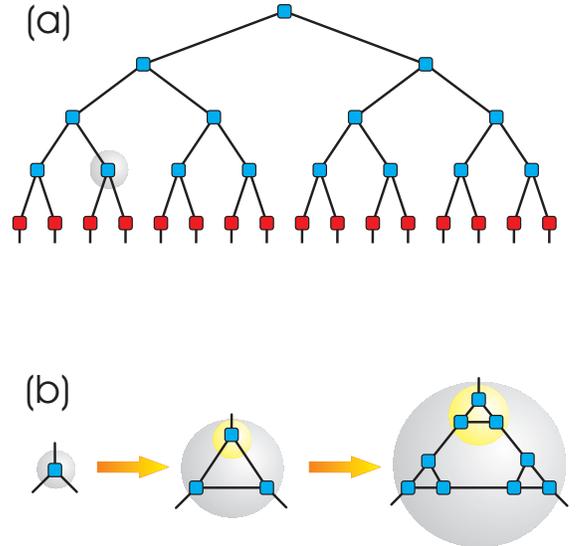}
\caption{(Color online). (a) Graphical representation of a tree tensor network (TTN). The boxes correspond to tensors, where joint indices are summed over. Open indices correspond to physical particles (red tensors), while auxiliary tensors are drawn in blue. The tensors are arranged in a tree-like structure, which guarantees that the contraction of the tensor network can be done efficiently. (b) Each of the tensors in the original tensor network is replaced by small triangular tensor structure in a concatenated fashion.}
\label{Fig_TTN}
\end{figure}

We consider now the example of a tree tensor network as shown in Fig.~\ref{Fig_TTN}a. Tree tensor networks are quasi-one dimensional structures that can -- similar to 1D chains or MPS -- be efficiently contracted in an exact way \cite{SDV06}. In our example, each of the tensors is of rank $3$, and has dimensions $d_1,d_2,d_3$, where $d_1=d$ for tensors with open indices. We replace each of the tensors by a small tensor network, which we choose to be a triangle. That is,
\be
A_{i_1i_2i_3}=\sum_{\alpha_1=1}^{D_1}\sum_{\alpha_2=1}^{D_2}\sum_{\alpha_3=1}^{D_3} B_{i_1 \alpha_3 \alpha_2}^1 B_{i_2 \alpha_3 \alpha_1}^2 B_{i_3 \alpha_1 \alpha_2}^3.
\ee
This process can now be iterated, i.e., each of the tensors $A_{i_1 \alpha_k \alpha_l}^i$ is replaced by three tensors, say $C_{\beta_1\beta_2\beta_3}^{i,j}$, in a triangular structure (see Fig.~\ref{Fig_TTN}b). There are two different types of tensors: External tensors -- i.e., ones which are connected to outside initial tensors -- of dimensions $d_i,D_j,D_k$ respectively, and internal tensors which have dimensions $D_i,D_j,D_k$.

We consider a situation where $D_i < d_j$. In such a case, the internal structure of the initial tensor $A_{i_1i_2i_3}$ is determined by elementary tensors $C_{\beta_1\beta_2\beta_3}^{i,j}$, and in general this restricts the values of $A_{i_1i_2i_3}$. Notice that the entanglement features of the corresponding CTS, as measured by the entropy of entanglement, are determined by the dimensions of the tensors, and are in particular limited by the dimension of the external links, i.e., $d_1,d_2,d_3$. That is, in terms of entanglement, nothing can be gained by introducing the internal tensor structure. In order that the resulting tensor network state can carry the same amount of entanglement as the one described by the initial tree tensor network, one needs that the dimension of inner links at concatenation level $k$ are larger than square root of the dimension of the links at concatenation level $k+1$. In particular, $D_1 \geq \sqrt{d_1}$ for $k=1$, while for $k=2$ tensor dimension (for the inner links) $D_2 \geq \sqrt{D_1} \geq d_1^{1/4}$ are required. This can easily be seen by considering bipartitions of the system and by noting that the achievable Schmidt rank is determined by the dimension and the number of links between the two groups.

The possible gain of such an internal tensor network structure is two-fold. First, the total number of parameters is reduced. While each initial tensor is described by $d_1d_2d_3$ parameters, the resulting tensor network of depth $k\geq 2$ is described by $(d_1+d_2+d_3)D^2 + 3^{k-2}D^3$ parameters, where we assumed $D_k=D$ for all internal links. Second, the size of each of the tensors in the internal tensor network structure is much smaller than the initial tensor. Many algorithms applied to the tensor network, e.g., the computation of normal forms of such tree tensor networks \cite{SDV06,VDB07,MS06}, or the optimization of tensors in a variational method \cite{HKHDVEP09}, scale with the dimension of the elementary tensors of the network. In spite of the usually polynomial scaling of these algorithms, the computations quickly become intractable for increasing $d_k$, so that a network containing tensors with small dimension are  favorable in general. We have utilized this approach in \cite{HKHDVEP09}, where numerical simulations using tree tensor networks are performed.

We remark that the contraction of the resulting tensor network becomes more difficult as compared to the initial tree structure. This is due to the fact that the concatenated tensor network contains loops. To retain numerical accessibility, either approximate treatments have to be applied (as in contraction schemes introduced in the context of PEPS \cite{MVC07}) or the tree-like structure has to be kept, e.g., by limiting the tree-width of the concatenated tensor network (as in Ref.~\cite{HKHDVEP09}).


\section{Applications}\label{sec:applications}

After having given some theoretical and analytical considerations for the possible advantages of CTS over other numerical methods for the description of states, we want to demonstrate applications of the CTS structure. The relevance of the CTS rests on two pillars. The first one is the ability of the (concatenated) tensor network to actually \emph{hold} the relevant information about a state. The analytical considerations above indicate that this is the case for states based on circuits, time evolved states and others. The second pillar is the question if we can, once given a CTS, read out the contained information. Progress has been made with very similar networks in the context of PEPS. What we want to demonstrate in the following is the ability to find the potentially good description with numerically accessible methods and see how good the approximation is. Moreover, this section has the aim to demonstrate the applicability of the known contraction methods and describe some improvements thereof.


\subsection{The descriptive potential of CTS}\label{sec:potential}

In this section, we want to demonstrate the descriptive potential of a CTS using a toy model. For reasons of comparison, relevant states of the toy model were calculated exactly and these exact states were then approximated with both MPS and CTS. To not infiltrate the CTS description with inaccuracies from an approximate read-out procedure, we used an exact contraction algorithm for this network~\footnote{We note that the scaling of this algorithm is exponential and hence not applicable to the case of larger networks.}.

In particular, we have tested the achievable accuracies when describing states resulting from time evolution in a spin chain, using the Hamiltonian
\be
\label{Isingtype}
H= \sum_{a} \sigma_z^{(a)}\sigma_z^{(a+1)}  + B \sum_a \sigma_x^{(a)},
\ee
with $B=1$ and a system size of $N=12$ physical sites. The system is initialized in the product state $|+\rangle^{\otimes N}$ and evolved over a time $T=3.5$, a point which is, in our units, close to the point where the fidelity of the CTS had a (periodically recurring) minimum. Time evolution under this Hamiltonian shows the typical growth of entanglement in the state that makes MPS-based description hard. The optimal tensors in the CTS and also the MPS description were approximated by optimizing the overlap of the exactly calculated state and the tensor network state in a sweeping procedure. For each tensor, the overlap
\[
\frac{|\langle\psi_{ex}|\psi_{CTS}\rangle|^2}{\langle\psi_{ex}|\psi_{ex}\rangle\langle\psi_{CTS}|\psi_{CTS}\rangle},
\]
was calculated, leaving out one tensor to optimize. This tensor can then be found using linear algebra techniques using the contraction result as a linear form. See Ref.~\cite{VC06b,MVC07} and Appendix~\ref{sec:variationalmethod}.

\begin{figure}[ht]
\includegraphics[width=.45\textwidth]{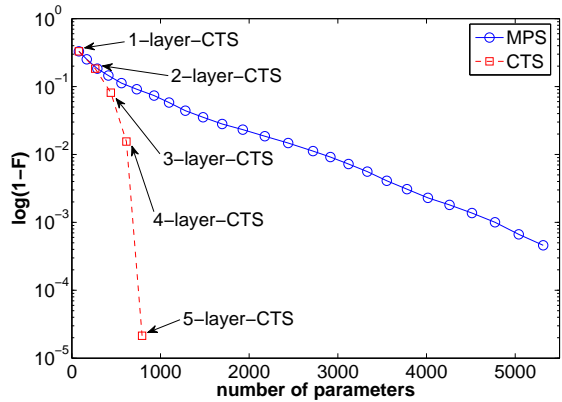}
\caption{(Color online). Comparison of the achievable accuracies when describing a time evolved state using MPS (blue, circles) and 2D CTS (red, squares). Simulated is the time evolution of a chain of $N=12$ physical sites, initialized in the product state $|+\rangle^{\otimes N}$ and evolved over a time $T=3.5$ using the Ising-Hamiltonian, Eq.~(\ref{Isingtype}), with $B=1$. After this time $T$ the chain is, in our units, close to the point where the fidelity of the CTS had a (periodically recurring) minimum. We compared MPS of varying dimension $\chi$, and 2D CTS with varying numbers of rows of auxiliary tensors and tensor dimensions, leading to the different number of parameters. Shown in the plot is the value $\log(1-F)$, where $F$ is the fidelity of the approximating state with the exact solution. Redundant parameters, as occuring in the boundary regions of the MPS, were not counted. We observe that the description in terms of a CTS is more efficient, i.e., a larger accuracy can be achieved when using the same number of parameters.}
\label{accuracy}
\end{figure}

We compare the achievable accuracy when describing the state with MPS of varying dimension $\chi$, and 2D CTS with varying numbers of rows of auxiliary tensors and tensor dimensions i.e., different $M$ and $D_k$. These variations lead to the different number of parameters that the comparison is based on. Although a quadratic growth of the parameter count is expected for the MPS using this method, the plot shows a approximately linear growth. This is due to the fact that we did not count redundant parameters, which occur in the matrices close to the boundaries of the chain. We observe (see Fig.~\ref{accuracy}) that the description in terms of a CTS is more efficient, i.e., both a larger accuracy can be achieved when using the same number of parameters, and for a fixed number of parameters one can describe the time evolution accurately for longer times using CTS.

Our tentative conclusion is that the additional structure leading to a reduction of the number of parameters and being imposed by the choice of CTS network reflects an internal structure to be found in the time evolved state itself, comparable to the Suzuki-Trotter expanded time evolution operator that can be programmed into the CTS. This is supported by the interpretation of the rows in the network to be operators acting on an MPS (the very first row of tensors). It seems natural to assume that the rows contain a version of the time evolution operator of the system. However, these operators have not been programmed into the network this time, but found by the optimization algorithm alone. Further investigations are necessary, but the idea that an optimization algorithm together with a suitable topological choice of network description yields a network of appropriate preparatory operators, reflecting deep structural properties of the described state, seems appealing.

\subsection{Reading out information from CTS}

So far we have only considered the possibility to efficiently {\em describe} quantum states in terms of concatenated tensor networks, but not how to efficiently {\em extract information} from such a description or how to update it. Both the extraction of information and also updating procedures rely on the \emph{contraction} of the tensor network, which is normally used in a slightly modified version for this purpose.

Given the fact that already a 2D tensor network is sufficient to describe all states resulting from a polynomially sized quantum computation, one does not expect an efficient contraction of such a tensor network to be possible in general. In fact, it has been shown in \cite{SWVC07} that contracting 2D tensor networks is computationally hard, the corresponding complexity class is $\#P$. However, this does not mean that no efficient {\em approximate} methods can exist which can successfully be applied in practice. In fact, in Ref.~\cite{VC06b} an approximate method to contract 2D tensor networks has been introduced and successfully applied in, e.g., the context of a ground state approximation for strongly correlated 2D systems \cite{MVC07}. This method will be described in the following, together with an investigation of two additional techniques: (i) A novel error correction scheme and (ii) An MPO compression scheme.

For specific tensor network topologies (e.g., networks corresponding to 1D graphs, trees or networks with a bounded tree width \cite{VDB07}), an exact and efficient contraction and update of tensors is possible. If we are free to choose the contraction order of all indices, there exist Monte-Carlo based methods for the contraction, whose application will be shown in Appendix B.


\subsubsection{Approximate contraction of 2D tensor networks}

The approximate contraction of a 2D tensor network with open boundary conditions, as introduced in Ref.~\cite{VC06b}, works as follows. The first (e.g., horizontal) line of tensors at the boundary can be interpreted as an MPS, where the lower indices are considered open. The second line can be viewed as a {\em matrix product operator} (MPO) acting on the first matrix product state. The resulting state (after contracting two lines) can again be described by an MPS, but of increased dimension. The aim is now to find (e.g., via a variational method) the optimal approximation of the resulting state by an MPS of {\em fixed} (low) dimension. This is, e.g., done by optimizing the individual tensors via solving a generalized eigenvalue problem (see Ref.~\cite{VC06b} or Appendix~\ref{sec:variationalmethod} below). The MPS found this way is now processed further, i.e., the MPO corresponding to the third line of tensors is applied, and one again aims at obtaining a proper approximation of the resulting state by an MPS of fixed dimension. The process is repeated until the second to last line of tensors is reached. The final step corresponds to calculating the overlap of the MPS resulting from above procedure (after processing all but the final line), and the MPS corresponding to the final line. All of these steps can be done efficiently. The evaluation of expectation values of (tensor product) observables works in a similar way. For details of the method, we refer the reader to \cite{MVC07}. Notice that the {\em same} method can be used for 2D tensor networks where some of the tensors are auxiliary tensors (without open indices), as we consider in this paper.

When using a concatenated MPS as described in Sec.~\ref{MPS_concat}, one may use the approximate method described above. However, especially when considering the description of time evolution (Sec.~\ref{MPS_time}) or (post-selected) quantum circuits (Sec.~\ref{MPS_circuit}), it is important to apply the method in a proper way, possibly utilizing symmetries of the state. In particular, in these cases the contraction should be done in the direction perpendicular to the order of the physical sites (left to right or right to left), rather than in lines parallel to the physical sites (up-down or down-up). A contraction in up-down direction would in these cases actually correspond to describing the state after each time step in terms of a fixed-sized MPS, and is actually equivalent to time evolution of an MPS as considered, e.g., in Ref.~\cite{DKSV04}. When using a contraction in the perpendicular direction, such a limitation does not apply, see also~\cite{CBprep}. Numerical evidence suggests a significant increase in accuracy in this case. Moreover, the treatment of infinitely extended, translationally invariant states leads to the observation that a contraction over infinitely many columns of tensors perpendicular to the physical direction often results in a projection onto the eigenspace with the largest-magnitude eigenvalues of the MPO represented by the column. This makes it possible to employ additional exact numerical techniques, see, e.g., Ref.~\cite{CBprep}.

If one is, like in the case of CTS, moreover able to choose the indices to contract freely, certain choices of tensors may allow for an efficient approximation via Monte-Carlo sampling techniques \cite{SV07}, see Appendix B. There, the application of Monte-Carlo methods to a 2D CTS will be demonstrated, using the inherent matrix product operator structure of the CTS. We would also like to mention the possibility to utilize String-bond state like tensor networks \cite{SWV07} in the context of CTS. Additionally, for certain choices of the tensors it is known that an exact and efficient contraction is possible \cite{MGcomb}.

In the following we would like to suggest two improvements for the traditional contraction scheme.


\subsubsection{An error correction scheme}

We will now describe an error-correcting procedure for the contraction of 2D CTS, which is applicable also to the contraction of other rectangular grids including PEPS.

We start with the traditional approximate contraction using the method described above, resulting in a number $\tilde{C}$, holding the contraction result. Following the line of argument from the sections above, we can interpret the number $\tilde{C}$ as an approximation of the number
\[
C=\langle M_1| M_2 \ldots M_{N-1}|M_N\rangle
\]
where $\langle M_1|$ is the MPS defined by the leftmost column of tensors, the operators $M_i$ are the MPO defined by the columns in the middle and $|M_N\rangle$ is the MPS defined by the rightmost column of tensors in the CTS. To remind the reader, a left to right contraction of the CTS involves the iteration of the following steps: (i) Start with $i=1$ and set $\langle\tilde{M}_1| := \langle M_1|$. (ii) Apply the MPO $M_{i+1}$ to the intermediate MPS $\langle\tilde{M}_{1,\cdots,i}|$. Both having a small bond-dimension, we obtain an MPS of large bond-dimension, $\langle M_{1,\cdots,i+1}|$. (iii) Reduce the bond-dimension of $\langle M_{1,\cdots,i+1}|$ to obtain another intermediate MPS $\langle \tilde{M}_{1,\cdots,i+1}|$, representing $\langle M_{1,\cdots,i+1}|$ as good as possible with this smaller bond-dimension. (iv) Increase $i$ by one and continue with step (ii). The aim of the error correcting scheme is to estimate the error introduced by cutting off the bond-dimension of the intermediate matrix product states, and to correct the result $\tilde{C}$ accordingly.

More precisely, after the $(i-1)$th step of the standard left to right contraction, the CTS is approximated by
\[
C  \approx \tilde{C}_{i-1} = \langle\tilde{M}_{1,\cdots,i}| M_{i+1} \cdots M_{N-1} |M_N\rangle
\]
where $\langle\tilde{M}_{1,\cdots,i}|\approx\langle M_1|M_2  \cdots M_i$. In the $i$th step we use the approximation $\langle\tilde{M}_{1,\cdots,i+1}| \approx \langle\tilde{M}_{1,\cdots,i}| M_{i+1}$ resulting in \be
C  \approx \tilde{C}_{i} = \langle\tilde{M}_{1,\cdots,i+1}| M_{i+2} \cdots M_{N-1} |M_N\rangle \notag
\ee
The additional error of $C$ in the $i$th approximation step is given by the value $\epsilon_i = \tilde{C}_{i-1}-\tilde{C}_i$, and the optimally corrected value of the contraction result is given by
\be
C=\tilde{C}+\sum_i \epsilon_i. \label{Ccorrection}
\ee

However, usually neither $\tilde{C}_{i-1}$ nor $\tilde{C}_i$ can be calculated exactly since the exact MPS description of the state $M_{i+2} \cdots M_{N-1} |M_N\rangle$ is too large to be computed. The crucial observation now is that the $(N-i-2)$th step of a \emph{right to left} contraction is a good approximation of this state with
\be
|\tilde{M}_{i+2,\cdots,N}\rangle \approx M_{i+2} \cdots M_{N-1} |M_N\rangle, \notag
\ee
which can be used to estimate the error $\epsilon_i$ produced by the $i$th step of the left to right contraction
\be
\epsilon_i &=& C_{i-1}-C_i \label{difference} \\
&\approx& \langle\tilde{M}_{1,\cdots,i}| M_{i+1}|\tilde{M}_{i+2,\cdots,N}\rangle \notag \\
&&-\langle{\tilde{M}_{1,\cdots,i+1}} |\tilde{M}_{i+2,\cdots,N}\rangle. \notag
\ee
This approximate value of $\epsilon_i$ is then used in Eq.~\ref{Ccorrection}.

For an estimation of the achievable accuracy with this error correction scheme, let the error of the overall left to right contraction be $\epsilon$. We note that also the error of the right to left contraction and its intermediate results $\langle\tilde{M}_{i,\cdots,n}|$ are of this size. Since the magnitude of the difference in Eq.~\ref{difference} is also of the order $\epsilon$, we are left with a residual absolute error of the order $\epsilon^2$ after the error correction. An application to toy models has confirmed our error estimation and yields a reduction of the error of about one order of $\epsilon$, or even better. For instance, the approximate contraction of the time evolved state in Sec.~\ref{sec:potential} with a cut-off bond-dimension $D=12$ results in a value $\tilde{C}$ with an error of $1.6(7)\%$ with and $25(10)\%$ without error correction, taking the mean of several approximations.

While a similar reduction could in principle be achieved by using a bigger cut-off bond-dimension for the intermediate results, the error correction scheme is favorable in most cases because of its better performance. As we can obtain all the required states $\langle \tilde{M}_{i,\cdots,n}|$ by caching one right to left contraction, we need merely twice the computation time for reducing the error by a factor of $\epsilon$. The overhead in memory depends on the cut-off bond-dimension of the states $\langle\tilde{M}_{i,\cdots,n}|$. Choosing this dimension equal to the dimension of the MPO $M_{i}$, the overhead is less than a factor of two, as we have to store $N-3$ extra MPS which is less than the $(N-2) \rm{MPO} + 2 \rm{MPS}$ of the CTS.

We remark that the applicability of this error correction scheme is not restricted to CTS, but can in a similar way also be used e.g. in the context of the 2D PEPS approach.


\subsubsection{Compressibility of sequences of matrix product operators}

Additionally, the number of tensors in the CTS description can be reduced significantly below the number needed in the canonical implementation of the Suzuki-Trotter picture, as given in section~\ref{MPS_time}, or for a generic network of (sparse) operators, like circuits.

The reason is that it is not necessary to restrict each row to the description of a single Suzuki-Trotter (or generic operator circuit) time step only. Instead we can first put a good approximation for many of these rows, applied successively, into \emph{one} row, thus using the descriptive power of the CTS to the maximal extend. This is possible by calculating and optimizing the overlap of one row of (variable) tensors with several concatenated rows of fixed tensors, in a way similar to maximizing the norm of a CTS when keeping every row but one fixed. We then concatenate these optimal rows, being fewer apparently, to reduce computational time in the read out process, whose computation time relies on the number of tensors involved.

To get an idea of the potential of this ansatz let us consider time evolution. When performing time evolution by a Suzuki-Trotter expansion with MPO compression, there are two possible sources of error. The first kind of errors comes from the MPO approximation. This kind of error can be controlled, as we know the fidelity of the replacement step (the overlap of the rows to be compressed with the replacement row). When this fidelity is too small, we can reduce the number of rows to be compressed. The second kind of error comes from the Suzuki-Trotter expansion itself. This error can be made small by choosing a very small time step, so that the MPO corresponding to one row is close to unity. After compressing two rows to one, we are able to iterate the compression and compress two already compressed rows to one row, which now represents four time steps. This way, we obtain one operator covering $2^n$ time steps with only $n$ compression steps. Taking initially very small time steps thus does not result in a big performance hit, as the compression is very strong here, i.e.\ exponential, and compensates for it. The compression can of course only be applied as long as one row can in principle hold the whole time step, but the results from the toy model in Sec.\ \ref{sec:potential} make us optimistic that a row has the potential to hold comparatively big time steps. Numerical evidence concerning the compression fidelity for a variety of operators supports this view. In this case, the compression not only improves the computation time but also the achievable precision by reducing the error introduced by Trotterization.

The variational ansatz just shown leads to a picture of the CTS where a fixed number of tensors will be employed in an optimal way, as opposed to the direct programming of a set of analytically accessible operators into the network.


\subsubsection{Special cases -- exact contraction}

Even though the problem of contracting an arbitrary 2D tensor network is in general computationally hard (\#P hard), under certain conditions an {\em efficient} and {\em exact} contraction of certain networks is possible.

One such example are planar tensor networks, where each of the tensors fulfills a so-called match-gate (or free fermion) condition \cite{MGcomb}. It follows that if we restrict ourselves to CTS corresponding to planar structures, one can calculate the norm as well as expectation values of tensor product of observables efficiently for such states, as long as all tensors in the tensor network fulfill the match-gate condition. This implies that one may use such CTS, e.g., as variational ansatz states for ground states or time evolution. In particular, we point out that the usage of auxiliary tensors as we propose for 2D CTS can be handled in exactly the same fashion.

Another example are networks corresponding to trees or structures with a bounded or only logarithmically growing tree width. These also can be contracted efficiently and exactly. For instance, the contraction of a subcubic tensor tree (i.e., a tensor network where each of the tensors in the tree is connected with three or less neighbors) has a computational effort scaling as $O(D^3)$. For a variational method for the search for ground states and the maximization of overlaps with CTS enhanced tree tensor networks, see Appendix A.


\subsubsection{The advantage of CTS in the operator picture}

Being able to program the Suzuki-Trotter expanded time evolution operator or other generic quantum circuits directly into the state description offers access to alternative advantageous numerical approaches. Time evolution methods usually rely on maximizing a significant number of overlaps of the kind
\[
\frac{|\langle\psi_{t+\delta t}| \tau (\delta t) |\psi_t\rangle|^2}{\langle\psi_{t+\delta t}|\psi_{t+\delta t}\rangle\langle\psi_{t}|\psi_{t}\rangle}
\]
where $\tau (\delta t)$ is the time evolution operator for a time $\delta t$, $|\psi_t\rangle$ is a known tensor network state and the tensors describing the state $|\psi_{t+\delta t}\rangle$ have to be found. In the context of circuits, an application of a set of gates can be regarded as a time step like above. Starting from this expression, to compute the time evolved state after a time $T$, one would have to compute single time steps repeatedly, and each time one would have to perform network contractions to determine the optimal tensor entries. (Algorithms of this kind are found to converge to a reasonable approximation of the best tensor network description of the desired time evolved state $|\psi_{T}\rangle$). Depending on the implementation, finding the optimal tensors can consist of many sub-steps, e.g., a sweeping procedure approximating single tensors while leaving the remaining tensors fixed, each sub-step requiring another contraction.

The CTS description of states is very efficient in this regard if, as given above, the time evolution operator $\tau(T)=\prod \tau(\delta t)$ is programmed into the structure and description of the state itself. Using CTS, we are able to circumvent the many contractions and possible sweeping steps by a single \emph{optimized} contraction. In contrast to a traditional MPS time evolution, for example, it is possible to use error correction and alternative contraction order (e.g., left-to-right instead of top-to-bottom). Moreover, we are able to employ MPO compression, which is, however, also applicable to the traditional time evolution method, but there not in a direction-optimized fashion.


\section{Summary and outlook}\label{sec:summary}

We have introduced concatenated tensor network states (CTS), a class of states that is obtained by decomposing the high-rank tensor describing the coefficient of a multi-particle states into a tensor network in an iterative fashion. The basic idea is to impose additional structure to each of the tensors appearing in a tensor network description of a given state.  We have demonstrated this approach for 1D tensor networks, where in a first step a description in terms of a matrix product state is obtained. Each of the matrices (tensors) is then further decomposed into a 1D tensor network (matrix product operator), yielding a 2D tensor network with many auxiliary tensors in the next step. Similar methods can be applied to 2D systems, yielding 2D or 3D PEPS with auxiliary tensors, or to tree tensor networks.

We have demonstrated that with such CTS, one can describe multi-particle quantum systems with rich entanglement features in an efficient way. In particular, states arising from time evolution or generated by polynomial (post-selected) quantum circuits can be described, and an interpretation in terms of (post selected) measurement-based quantum computation can be given. The states can -- in contrast to matrix product states or projected entangled pair states -- contain a large amount of block-wise entanglement (up to a volume law) and long-ranged correlations, while their description remains efficient. In particular, a subclass of matrix product states and projected entangled pair states with high effective bond dimension can be described.

We have demonstrated that it is possible to describe states arising from time evolution of a 1D quantum system with help of such a 2D CTS more efficiently than with a matrix product state. We have discussed the description arising from a Trotter decomposition of the evolution operator, as well as direct optimization of (auxiliary) tensors in the 2D tensor network of given size and dimension. In this context, we have also applied a method to compress matrix product operators to obtain a more efficient description of the time-evolved state.

We have also discussed and improved methods to read out information from 2D tensor networks. The applicability of approximate contraction methods, possibly with different direction of contraction (left to right), has been discussed and improved using an error correction scheme. Both the effectiveness of the CTS description in the context of time evolution of one-dimensional systems as well as the impact of our suggested enhancements to the traditional read-out methods were demonstrated using numerical results for a toy model. Also the applicability of Monte-Carlo methods for the contraction was demonstrated.

The results indicate that the new class of states is useful in the context of describing and simulating time evolution of 1D quantum systems, but might also be used for the simulation of ground states of 2D quantum systems. The different interpretations in terms of trotter decomposition, (post selected) quantum networks or (post selected) measurement-based quantum computation we provide may also inspire a new point of view to tensor network states and encourage further development.

\begin{acknowledgments}
We thank M. Van den Nest for interesting discussions. This work was supported by the FWF and the European Union (QICS, SCALA).
\end{acknowledgments}

{\em Note added}: We would like to point the reader to Ref. \cite{CBprep}, where methods similar to the one described in this paper have been independently derived and utilized in the context of time evolution in infinite systems.

\appendix


\section{Variational optimization of CTS-enhanced tree tensor networks}\label{sec:variationalmethod}

Important applications of quantum mechanical simulations are the search for ground states and the computation of the time evolution of states governed by a given Hamiltonian, usually employing variational methods. On the mathematical level, an essential element of the variational procedures in tensor networks is the linear dependence between the network and each of its tensors. Contracting the CTS (or an amplitude or expectation value involving a CTS) leaving out one of the tensors provides us with a simple linear or quadratic form which is suitable for investigation. The maximization of an overlap or minimization of an energy is thus reduced to the analysis of such a form and can be performed using linear algebra. Naturally a possibly exact and efficient contraction method is desired.

More precisely, for instance, finding the ground state of a Hamiltonian $H$ means finding the state $\ket{\psi}$ that solves
\[
\frac{\bra{\psi}H\ket{\psi}}{\braket{\psi}{\psi}}\overset{!}{=}\mbox{min}.
\]
We can write a tensor network state as
\begin{equation}
\ket{TNS}:=\sum_{\mathbf{s},\mathbf{a}}T_{s_{1}...s_{n}a_{1}...a_{n}}R_{s_{n+1}...s_{N}a_{1}...a_{n}}\ket{\mathbf{s}},\label{eq:TNSdef}
\end{equation}
where $T$ is the tensor under consideration and $R$ is the remainder of the tensor network, already contracted up to the indices that connect $T$ and $R$. The mentioned linear dependence on $T$ is exploited by (virtually) replacing the tensor $T$ by tensors $D\left(\tilde{\mathbf{s}},\tilde{\mathbf{a}}\right)$ which have the entries
\begin{equation}
D\left(\tilde{\mathbf{s}},\tilde{\mathbf{a}}\right)_{\mathbf{s},\mathbf{a}}:=\begin{cases}
1 & \mathbf{s}=\tilde{\mathbf{s}}\mbox{ and }\mathbf{a}=\tilde{\mathbf{a}}\\
0 & \mbox{else}\end{cases}.\label{eq:Ddef}
\end{equation}
With help of the tensors $D$ we generate states
\[
\ket{\psi\left(\tilde{\mathbf{s}},\tilde{\mathbf{a}}\right)}:=\sum_{\mathbf{s},\mathbf{a}}D\left(\tilde{\mathbf{s}},\tilde{\mathbf{a}}\right)_{s_{1}...s_{n}a_{1}...a_{n}}R_{s_{n+1}...s_{N}a_{1}...a_{n}}\ket{\mathbf{s}}
\]
where $\left(\tilde{\mathbf{s}},\tilde{\mathbf{a}}\right)$ is a combined index. With these states we in turn generate matrices
\[
E_{\left(\tilde{\mathbf{s}},\tilde{\mathbf{a}}\right),\left(\tilde{\mathbf{s}}',\tilde{\mathbf{a}}'\right)}:=\braket{\psi\left(\tilde{\mathbf{s}},\tilde{\mathbf{a}}\right)}{H|\psi\left(\tilde{\mathbf{s}'},\tilde{\mathbf{a}}'\right)}\]
as well as\[
N_{\left(\tilde{\mathbf{s}},\tilde{\mathbf{a}}\right),\left(\tilde{\mathbf{s}}',\tilde{\mathbf{a}}'\right)}:=\braket{\psi\left(\tilde{\mathbf{s}},\tilde{\mathbf{a}}\right)}{\psi\left(\tilde{\mathbf{s}'},\tilde{\mathbf{a}}'\right)}.\]
Finding the entries of the tensor $T$ is now reduced to a generalized eigenvalue problem,
\[
\frac{t^{*} \cdot E \cdot t}{t^{*} \cdot N \cdot t}\overset{!}{=}\mbox{min},
\]
where the tensor $t$ with the minimum generalized eigenvalue
\[
E \cdot t = \lambda N \cdot t
\]
is the solution of the local minimization problem, i.e., the minimization problem with respect to $T$ when the other tensors are fixed. The exact technical implementation of this idea is, of course, subject to optimization and will not be done by the mentioned contraction over dummy tensors.

We are facing two numerical problems. One is the contraction of the tensor network. The second problem is finding the generalized eigenvalues of the matrices given in the section above. The difficulty of the contraction of a network increases polynomially with the index rank and moreover depends strongly on the topological structure of the network. Tree networks can be contracted with efficient algorithms \cite{SDV06}, and the dimension of the tensors is the parameter which governs the efficiency of the contraction in this case. Contracting networks with loops on the other hand is in general intractable \emph{if performed exactly}.

Obviously, the nested (e.g, triangular) tensor structure does not simplify the contraction of the network, but remains feasible if the tree-width of the tensor network is small. A tree whose tensors are replaced by small loops is such a tree-like structure. In the corresponding efficiency considerations, the role of the dimension connecting the tensors is replaced by the dimension connecting \emph{the loops among each other}. Now, there are values of the dimension (entering polynomially into the computational effort) where a contraction of the network is still possible, but a solution of the generalized eigenvalue problem is not -- the reason being the size of the corresponding matrices, whose size is scaling like $D^{3}\times D^{3}$ if $D$ is the index rank.

The nested tensor loops address this problem by ``shielding'' the large outgoing index. More precisely, let us define a tensor network state like in Eq.~\ref{eq:TNSdef}. For the sake of simplicity we consider a tensor $A$ \emph{not} connected to any physical sites and the network to be a subcubic tree, \[ \ket{TNS}:=\sum_{\mathbf{s},\mathbf{a}}A_{a_{1}a_{2}a_{3}}R_{s_{n+1}...s_{N}a_{1}a_{2}a_{3}}\ket{\mathbf{s}}.\] We are now able to rewrite the state $\ket{TNS}$ by replacing $A$ by a loop as shown in Fig.~\ref{Fig_TTN}
\[
A_{a_{1}a_{2}a_{3}}:=\sum_{\alpha\beta\gamma}B_{a_{1}\alpha\beta}^{1}B_{a_{2}\alpha\gamma}^{2}B_{a_{3}\beta\gamma}^{3}.
\]
where now correspondingly
\begin{multline*}
\ket{TNS_{\mbox{\tiny{loop}}}}\\
:=\sum_{\mathbf{s},\mathbf{a}}\sum_{\alpha\beta\gamma}B_{a_{1}\alpha\beta}^{1}B_{a_{2}\alpha\gamma}^{2}B_{a_{3}\beta\gamma}^{3}R_{s_{n+1}...s_{N}a_{1}a_{2}a_{3}}\ket{\mathbf{s}},
\end{multline*}
This helps to reduce the size of the matrix of the corresponding eigenvalue problem because a) the tensors $B^{1},B^{2},B^{3}$ can be locally optimized individually, and b) the indices $\alpha,\beta,\gamma$ can have smaller dimension, while \emph{the loop structure of the tensor $T$ replacement network retains the entanglement properties} which are so important for the power of the description. It is possible to choose low but sufficiently high index rank for the internal indices such that the entanglement being carried by the external indices (that connect the loops among each other) is not reduced.

In detail, finding the optimal values of the loop tensors $B^{i}$ can be performed as follows. First, the network represented by $R$ has to be contracted. Once this tensor is found, it is kept fixed for the optimization of the tensors $B^{i}$. We then repeat the optimization steps for the loop tensors as described in the section above, using the state
\begin{multline*}
\ket{TNS_{\mbox{\tiny{loop}}}}
:=\sum_{\mathbf{s},\mathbf{a}}\sum_{\alpha\beta\gamma}d^{1}\left(\tilde{a}_{1},\tilde{\alpha},\tilde{\beta}\right)_{a_{1}\alpha\beta}B_{a_{2}\alpha\gamma}^{2}B_{a_{3}\beta\gamma}^{3}\\
\times R_{s_{n+1}...s_{N}a_{1}a_{2}a_{3}}\ket{\mathbf{s}},
\end{multline*}
with a tensor $d^{1}$ like in Eq.~\ref{eq:Ddef}. Similarly, we proceed for the tensors $B^{2},B^{3}$. In these steps we can make use of the fact that several (more than one) sweeps through the loop tensors will give a better convergence, while the computational overhead for this is small, because the huge remainder of the network -- represented by the tensor $R$ -- stays constant and does not need to be contracted again. If the dimension of the internal indices is large enough, several sweeps through the loop will converge to a network
\[ A_{a_{1}a_{2}a_{3}}:=\sum_{\alpha\beta\gamma}B_{a_{1}\alpha\beta}^{1}B_{a_{2}\alpha\gamma}^{2}B_{a_{3}\beta\gamma}^{3},
\]
with a tensor $A$ whose values are the same as in the case without the replacement network. In some cases the original problem of finding $A$ would not have been feasible, but even if so, the sweeping through the loop gives an advantage in computation time.


\section{Monte-Carlo sampling of CTS}\label{sec:montecarlo}

In some instances, it is possible to contract the concatenated tensor network approximately with a Monte-Carlo based approach. For this, let us quickly recall how the Monte-Carlo method works. The easiest and most basic Monte-Carlo (MC) technique is the Metropolis algorithm \cite{Metropolis}. Like all MC methods, it is used to estimate integrals (or sums) over high dimensional integration spaces. In these spaces, naïve approaches like Riemann-integration require a huge number of sampling points for a certain required accuracy, whereas usually the MC methods show a much quicker convergence to the exact value.

The basic idea is that we can select a sample of points in the integration space such that
\[
Z^{-1}\int_{V}f\left(x\right)\mu\left(x\right)dx\approx N^{-1}\sum_{\left\{ x_{i}\right\} _{i=1}^{N}\subset V}f\left(x_{i}\right)\triangle v,
\]
where $\triangle v$ is a unit volume in $V$, $\mu$ is a well-behaved measure on $V$, and $Z=\int_{V}\mu\left(x\right)dx$. Naturally, the selection rule for the set of samples, $\left\{ x_{i}\right\} $, is the key and has a foundation in statistical mechanics. Assume that $f$ is a property of an ergodic physical system with density (probability density to be found at that point) $\mu$ in configuration space. The system being ergodic, we know that the time average of the property $f$ equals the average of $f$ over configuration space with weight $\mu$,
\[
\left\langle f\right\rangle _{t}=\int_{V}f\left(x\right)\mu\left(x\right)dx.
\]

We obtain the set of samples $\left\{ x_{i}\right\} $ by simulating the behavior of the system in time and recording the position $x\left(t_{i}\right)=x_{i}$ at discretely (and equally) spaced points $\left\{ t_{i}\right\} $ in time. Let now $P\left(x\rightarrow x'\right)$ be the probability of the system to go, during one discrete time step of a random walk, from point $x$ to point $x'$. A set $\left\{ x_{i}\right\} $ of a random walk derived with such a rule is called a \emph{Markov chain}, with the essential property being that $x_{i}$ is only dependent on $x_{i-1}$ (and not $x_{i-2}$ etc.). It is known that the so called \emph{detailed balance condition} for the probability $P$,
\[
\mu\left(x\right)P\left(x\rightarrow x'\right)=\mu\left(x'\right)P\left(x'\rightarrow x\right),
\]
is a sufficient criterion to ensure that a random walk of the system, ruled by the transfer probability $P$, yields a time average approaching the value $Z^{-1}\int_{V}f\left(x\right)\mu\left(x\right)dx$ for $t\rightarrow\infty$. One can impose this transfer probability by the following rule:

\begin{enumerate}
\item Being at point $x_{i}$, choose randomly a position $\xi$.
\item Calculate the value $A\left(x_{i}\rightarrow\xi\right)=\mbox{min}\left(1,\frac{\mu\left(\xi\right)}{\mu\left(x_{i}\right)}\right).$
\item Randomize a number in the interval $a\in\left[0,1\right]$.
\item If $a<A\left(x_{i}\rightarrow\xi\right)$, then $x_{i+1}=\xi$, otherwise
$x_{i+1}=x_{i}$.
\end{enumerate}

This is the (basic) Metropolis algorithm \cite{Metropolis}. The probability of going from $x$ to $x'$ under this algorithm obeys the detailed balance condition and hence yields a sample that is representative for the measure $\mu$. We note that with this rule we can generate arbitrarily large sets of positions in time without the need to store the set $\left\{ x_{i}\right\} $ itself. Furthermore it is not necessary (for the evaluation of the time evolution) to know the value $Z=\int_{V}\mu\left(x\right)dx$, which cancels in the calculation of $A$; a fact that makes it possible to work with relative probabilities and unnormalized measures.

We now want to show that the contraction of the concatenated tensor networks can be implemented via a MC algorithm. To demonstrate the principle we give the formulas to contract a toroidal network of $N\times M$ tensors of rank $4$, although the formalism is easily adapted to non-toroidal networks and higher dimensions. Consequently, we want to calculate
\[
\sum_{\mbox{indices }s_{u},s_{d},s_{l},s_{r}}\prod_{i,j}T_{s_{u}\left(i,j\right)s_{d}\left(i,j\right)s_{l}\left(i,j\right)s_{r}\left(i,j\right)}^{i,j}
\]
where $u,d,l,r$ mean ``up, down, left, right'' respectively, the indices $s_{u,d,l,r}$ depend on the position $\left(i,j\right)$ within the network, and $s_{d}\left(i,j\right)=s_{u}\left(i+1,j\right)$, $s_{r}\left(i,j\right)=s_{l}\left(i,j+1\right)$, $s_{u}\left(1,j\right)=s_{d}\left(N,j\right)$ and $s_{l}\left(i,1\right)=s_{r}\left(i,N\right)$.

The basic principle is to perform the contraction over the indices $s_{u,d,l,r}$ in a hierarchical order: We first contract over the indices in each row. This is formally the trace over a product of matrices, the matrices being the tensors of rank four, where the indices connecting in the vertical direction are kept fixed. In the next step, we contract over the indices that connect the rows. Following this idea, in the case of an $n$-dimensional network, the hierarchy has $n$ levels -- indices of increasing level thereby connecting slices of increasing dimensionality. For the problem at hand, we write
\begin{multline}
\sum_{\begin{array}{c}
s_{u},s_{l}\end{array}}\prod_{i,j}T_{s_{u}\left(i,j\right)s_{u}\left(i+1,j\right)s_{l}\left(i,j\right)s_{l}\left(i,j+1\right)}^{i,j}\\
=\sum_{s_{u}}\prod_{i}R_{\mathbf{s}_{u}\left(i\right)\mathbf{s}_{u}\left(i+1\right)}^{i},\label{eq:rowproduct}
\end{multline}
where $\mathbf{s}_{u}\left(i\right)=\left(s_{u}\left(i,1\right),s_{u}\left(i,2\right),...\right)$,
\begin{multline*}
\prod_{i}R_{\mathbf{s}_{u}\left(i\right)\mathbf{s}_{u}\left(i+1\right)}^{i}:=\sum_{s_{l}}\prod_{j}T_{s_{u}\left(i,j\right)s_{u}\left(i+1,j\right)s_{l}\left(i,j\right)s_{l}\left(i,j+1\right)}^{i,j}\\
=\mbox{Tr}\left[\prod_{j}T^{i,j}\left[s_{u}\left(i,j\right),s_{u}\left(i+1,j\right)\right]\right],\end{multline*}
and where $T^{i,j}\left[\cdot,\cdot\cdot\right]$ are matrices with elements
\[
\left(T^{i,j}\left[a,b\right]\right)_{c,d}:=T_{a,b,c,d}^{i,j}.
\]

What we see is that, while keeping the \emph{up} and \emph{down }indices fixed in each row, we obtain a set of matrices for each row, and carry out the summation of the \emph{left }and \emph{right} indices with matrix products and a trace. As seen in Eq.~\ref{eq:rowproduct}, this leaves us with another set of matrices, $R_{\mathbf{s}_{u}\left(i\right)\mathbf{s}_{u}\left(i+1\right)}^{i}$, which are treated like the matrices $T^{i,j}\left[\cdot,\cdot\cdot\right]$: they are multiplied and the product is traced over. This means that the contraction over the whole set of indices is the trace over a matrix product for matrices whose elements are again traces over matrix products. This recursion repeats itself and adds another ``generation'' of matrix products for every dimension of the tensor network to be contracted. If the tensor network was no torus, the boundaries would have been taken by tensors of lower rank. This would then replace the trace by the contraction with the tensor at the boundary; the situation is similar to matrix product states with periodic versus open boundary conditions.

As we see, formally, we use matrix products and traces over all levels of matrices in the hierarchy, but for the second generation of matrix products and higher ones this cannot be carried out explicitly anymore. While the dimensions of the $T^{i,j}\left[\cdot,\cdot\cdot\right]$-matrices depend on our arbitrary choice of the index rank of the tensors $T$ \emph{only} -- making the first generation of matrix products feasible -- the dimensions of the second generation of matrices (here: the row matrices $R$) is already much too large: The number of columns enters exponentially. This is the point where the Monte-Carlo sampling is used. We sample the (second and higher) generation of matrix products.

To improve readability, we leave out the index $u$ from now on and write the matrix elements in the following way $R_{\mathbf{s}_{u}\left(i\right)\mathbf{s}_{u}\left(i+1\right)}^{i}=\left\langle \mathbf{s}_{i}\right|R^{i}\left|\mathbf{s}_{i+1}\right\rangle $ (as usual in quantum mechanics). There are in principle two possible ways how to perform the MC-approach \emph{in detail.}

In the first, the contraction takes the form\begin{multline}
\sum_{\mathbf{s}}\prod_{i}R_{\mathbf{s}\left(i\right)\mathbf{s}\left(i+1\right)}^{i}\\
=\sum_{\mathbf{s}_{1},...,\mathbf{s}_{N+1}}\left\langle \mathbf{s}_{1}\right|R^{1}\left|\mathbf{s}_{2}\right\rangle \!\left\langle \mathbf{s}_{2}\right|R^{2}...R^{n}\left|\mathbf{s}_{n+1}\right\rangle \\
\times\left\langle \mathbf{s}_{n+1}\right|R^{n+1}\left|\mathbf{s}_{n+2}\right\rangle \left\langle \mathbf{s}_{n+2}\right|R^{n+2}...R^{N}\left|\mathbf{s}_{N+1}\right\rangle \\
=\sum_{\mathbf{s}_{1},...,\mathbf{s}_{N+1}}L\left(\mathbf{s}_{1},...,\mathbf{s}_{N+1}\right)R\left(\mathbf{s}_{1},...,\mathbf{s}_{N+1}\right)\\
=\sum_{\mathbf{s}_{1},...,\mathbf{s}_{N+1}}\left|L\left(\mathbf{s}_{1},...,\mathbf{s}_{n+1}\right)\right|^{2}\frac{R\left(\mathbf{s}_{n+1},...,\mathbf{s}_{N+1}\right)}{L\left(\mathbf{s}_{1},...,\mathbf{s}_{n+1}\right)^{*}}\\
=\sum_{\mathbf{s}_{1},...,\mathbf{s}_{N+1}}\mu\left(\mathbf{s}_{1},...\mathbf{s}_{n+1}\right)f\left(\mathbf{s}_{1},...,\mathbf{s}_{N+1}\right).\label{eq:MC-ansatz-one}\end{multline}
where
\begin{eqnarray*}
L\left(\mathbf{s}_{1},...,\mathbf{s}_{n+1}\right) & := & \left\langle \mathbf{s}_{1}\right|R^{1}\left|\mathbf{s}_{2}\right\rangle \!\left\langle \mathbf{s}_{2}\right|R^{2}...R^{n}\left|\mathbf{s}_{n+1}\right\rangle \\
R\left(\mathbf{s}_{n+1},...,\mathbf{s}_{N+1}\right) & := & \left\langle \mathbf{s}_{n+1}\right|R^{n+1}\left|\mathbf{s}_{n+2}\right\rangle \\
 &  & \quad\times\left\langle \mathbf{s}_{n+2}\right|R^{n+2}...R^{N}\left|\mathbf{s}_{N+1}\right\rangle,
\end{eqnarray*}
$L$ and $\mu$ are actually independent of $\mathbf{s}_{k},k>n+2$ and $R$ is independent of $\mathbf{s}_{k},k<n$. This way, we have derived a formal expression that is compatible with an MC-approach: $\mu$ is a positive measure (obviously, $\mu\geq0$), $f$ is a function to be integrated over and weighed by $\mu$. The integration (summation) space is the set of all vectors $\mathbf{S}=\left(\mathbf{s}_{1},...,\mathbf{s}_{N+1}\right)$.

The second, alternative, implementation of the MC-approach is similar, but divides the integration space into smaller subsections. We use the same ansatz, but this time, we do not look at the functions $L$ and $R$, but at the functions $l$ and $r$
\begin{eqnarray*}
l\left(\mathbf{s}_{1},\mathbf{s}_{n+1}\right) & := & \sum_{\mathbf{s}_{2},...,\mathbf{s}_{n}}L\left(\mathbf{s}_{1},...,\mathbf{s}_{n+1}\right)\\
r\left(\mathbf{s}_{n+1},\mathbf{s}_{N+1}\right) & := & \sum_{\mathbf{s}_{n+2},...,\mathbf{s}_{N}}R\left(\mathbf{s}_{n+1},...,\mathbf{s}_{N+1}\right)
\end{eqnarray*}
instead. Then
\begin{multline*}
\sum_{\mathbf{s}}\prod_{i}R_{\mathbf{s}\left(i\right)\mathbf{s}\left(i+1\right)}^{i}\\
=\sum_{\mathbf{s}_{1},\mathbf{s}_{n+1,}\mathbf{s}_{N+1}}\left|l\left(\mathbf{s}_{1},\mathbf{s}_{n+1}\right)\right|^{2}\frac{r\left(\mathbf{s}_{n+1},\mathbf{s}_{N+1}\right)}{l\left(\mathbf{s}_{1},\mathbf{s}_{n+1}\right)^{*}},
\end{multline*}
where the sum is performed over only three (instead of $N+1$) vectors. The functions $l$ and $r$, which are sums themselves, can be calculated, again, with an MC-approach. The expression
\[
l\left(\mathbf{s}_{1},\mathbf{s}_{n+1}\right)=\sum_{\mathbf{s}_{2},...,\mathbf{s}_{n}}\left\langle \mathbf{s}_{1}\right|R^{1}\left|\mathbf{s}_{2}\right\rangle \!\left\langle \mathbf{s}_{2}\right|R^{2}...R^{n}\left|\mathbf{s}_{n+1}\right\rangle
\]
(and likewise $r$) are very similar formally to the original expression for the whole sum in Eq.~\ref{eq:MC-ansatz-one} and can be treated in an analogous fashion: another subdivision of the summation space $\left\{ \mathbf{s}_{2},...,\mathbf{s}_{n}\right\} $ into two parts will be initiated. This recursion will have a depth of order $\log\left(N\right)$; hence, assuming we need for a summation over three vectors $O\left(m\right)$ MC-samples, this method will lead to a computational effort of $O\left(\left(2m\right)^{\log N}\right)=O\left(2^{\log N}N^{\log m}\right)$. The decision to choose way $1$ or way $2$ should be based on tests concerning the achieved convergence speed and depend on the problem at hand.

There is one more bookkeeping-issue to consider in both approaches. Let $\mathfrak{S}\left[\cdot\right]$ denote the limes of the Monte-Carlo sampling over a Markov chain $\left\{ \mathbf{S}_{i}\right\} $, approaching infinite length, that is generated by the measure $\mu$. Then
\[
\left(\sum_{\mbox{\tiny{all} }\mathbf{S}}\mu\left(\mathbf{S}\right)\right)\mathfrak{S}\left[f\right]=\sum_{\mbox{all }\mathbf{S}}\mu\left(\mathbf{S}\right)f\left(\mathbf{S}\right),
\]
that is, the Monte-Carlo summation yields a result as if the measure was \emph{normalized}. Hence we need an estimate for the partition sum $Z=\sum_{\mbox{all }\mathbf{S}}\mu\left(\mathbf{S}\right)$. It can be obtained through the relationship
\[
\mathfrak{S}\left[\mu^{-1}\right]=Z^{-1}\sum_{\mbox{all }\mathbf{S}}\mu\left(\mathbf{S}\right)\mu^{-1}\left(\mathbf{S}\right)=Z^{-1}\left|\left\{ \mathbf{S}\right\} \right|,
\]
hence $Z=\left|\left\{ \mathbf{S}\right\} \right|\left(\mathfrak{S}\left[\mu^{-1}\right]\right)^{-1}$. This doubles the computational effort.

The applicability of the MC method is not universal though, the limits being set by the measure $\mu$ and the function $f$. As mentioned in the description of the MC method, the system needs to be ergodic. This is not the case if the measure $\mu$ has regions whose borders cannot be crossed during a random walk. This phenomenon is usually due to the measure being zero (or very small compared to other regions) in some regions of the configuration space. In this case, the random walker decides too often to stay where it is and traverses the configuration space either too slowly or not at all, obviously not yielding an interesting sample collection. A property that can cause non-ergodicity is hence sparseness of the tensors in the concatenated tensor network. A remedy for this problem is a local change of basis, and hence readily available.

Another, more serious, problem is non-convergence of the MC sampling due to the so called sign-problem. This means that the relative error of the sampling method stays big over an exponentially large sampling set, because, roughly speaking, in the summation process two very big sub-summands (with small relative error) cancel each other. This problem does not occur if all (row-)tensor elements are $\geq0$.

Also other tensor structures are known to be accessible in similar ways. For instance, so called string-bond states \cite{SWV07} can be efficiently contracted via Monte-Carlo sampling. While in the initial work only string-bond states where all tensors are associated with physical particles have been considered,  one can easily adopt the construction of concatenated tensor network states to string-bond states. In particular, can use exactly the same tensor structures as for string-bond states, where however only some of the tensors correspond to physical particles, while the other tensors are auxiliary.

We have performed numerical tests on this method and found that for certain cases of tensor networks, e.g., with positive tensor entries, the algorithm converges quickly to a value of the contraction result with a quickly decreasing relative error.

\end{document}